\documentclass[twocolumn]{article}

\usepackage{arxiv}

\usepackage[utf8]{inputenc} 
\usepackage[T1]{fontenc}    
\usepackage{hyperref}       
\usepackage{url}            
\usepackage{booktabs}       
\usepackage{amsfonts}       
\usepackage{nicefrac}       
\usepackage{microtype}      
\usepackage{lipsum}
\usepackage{graphicx}
\usepackage{multicol}
\usepackage[font={small,it}]{caption}
\graphicspath{ {./images/} }

\usepackage{amsmath,amsfonts}
\usepackage{algorithmic}
\usepackage{algorithm}
\usepackage{array}
\usepackage{subfig}
\usepackage{textcomp}
\usepackage{stfloats}
\usepackage{url}
\usepackage{verbatim}
\usepackage{graphicx}
\usepackage{cite}
\usepackage{subcaption}
\usepackage{siunitx}
\usepackage[font={small,it}]{caption}
\usepackage[export]{adjustbox}

\title{Ultra-low latency quantum-inspired machine learning predictors implemented on FPGA}

\author{
  Lorenzo Borella \\
  Istituto Nazionale di Fisica Nucleare (INFN), Padova\\
  Dipartimendo di Fisica e Astronomia (DFA) \\
  University of Padua, IT
  \texttt{lorenzo.borella.1@phd.unipd.it} \\
   \And
  Alberto Coppi \\
  Dipartimendo di Fisica e Astronomia (DFA)\\
  University of Padua, IT
  \And
  Jacopo Pazzini \\
  Istituto Nazionale di Fisica Nucleare (INFN)\\
  Dipartimendo di Fisica e Astronomia (DFA)\\
  Dipartimendo di Ingegneria dell'Informazione (DEI)\\
  Dipartimendo di Ingegneria Industriale (DII)\\
  University of Padua, IT
   \And
   Andrea Stanco \\
   Istituto Nazionale di Fisica Nucleare (INFN)\\
   Padua Quantum Technology Research Center\\
   Dipartimendo di Ingegneria dell'Informazione (DEI)\\
   University of Padua, IT
 \And
   Marco Trenti \\
   Tensor AI Solutions GmbH, Pfaffenhofen a.d. Roth, Germany \\
 \And
   Andrea Triossi \\
   Istituto Nazionale di Fisica Nucleare (INFN) \\
   Dipartimendo di Fisica e Astronomia (DFA) \\
   University of Padua, IT
   \And
   Marco Zanetti \\
   Istituto Nazionale di Fisica Nucleare (INFN) \\
   Dipartimendo di Fisica e Astronomia (DFA) \\
   University of Padua, IT}

\begin{document}

\onecolumn 
\maketitle
\begin{abstract}
    Tensor Networks (TNs) are a computational paradigm used for representing quantum many-body systems. Recent works have shown how TNs can also be applied to perform Machine Learning (ML) tasks, yielding comparable results to standard supervised learning techniques.In this work, we study the use of Tree Tensor Networks (TTNs) in high-frequency real-time applications by exploiting the low-latency hardware of the Field-Programmable Gate Array (FPGA) technology. We present different implementations of TTN classifiers, capable of performing inference on classical ML datasets as well as on complex physics data. A preparatory analysis of bond dimensions and weight quantization is realized in the training phase, together with entanglement entropy and correlation measurements, that help setting the choice of the TTN architecture. The generated TTNs are then deployed on a hardware accelerator; using an FPGA integrated into a server, the inference of the TTN is completely offloaded. Eventually, a classifier for High Energy Physics (HEP) applications is implemented and executed fully pipelined with sub-microsecond latency. 
\end{abstract}


\twocolumn

\section{Introduction}
Tensor Network (TN) methods are commonly used to represent and simulate many-body quantum systems on classical computers~\cite{simulatingquantum,nature_parallelization,PhysRevResearch.5.013031}. They consist of the factorization of very high-order tensors into networks of smaller tensors, in this way avoiding the curse of dimensionality~\cite{Evenbly_2011}. Tree Tensor Networks (TTNs) are the most general loopless TN architecture,  originally devised to represent the wave functions of weakly entangled states and to study their evolution.

Besides their pure quantum applications, the properties of TTNs and their simple minimization algorithms can also be exploited to solve canonical Machine Learning (ML) tasks~\cite{chen2023machinelearningtreetensor,wang2023tensornetworksmeetneural,sengupta2022tensornetworksmachinelearning}. The inherent quantum characteristics of these networks provide them with valuable properties that enable us to gain insightful perspectives into the distribution of information within the network. For instance, entanglement entropy measurements can be performed on the links of the TTN, retrieving a quantitative estimation of the relevance of the information stored in each node~\cite{10.1093/ptep/ptad018}. Moreover, it is straightforward to measure the quantum correlations between data features, eventually eliminating redundancies and reducing the number of active parameters in the overall architecture. 

\begin{figure}[h]
    \centering
    \includegraphics[width=\linewidth]{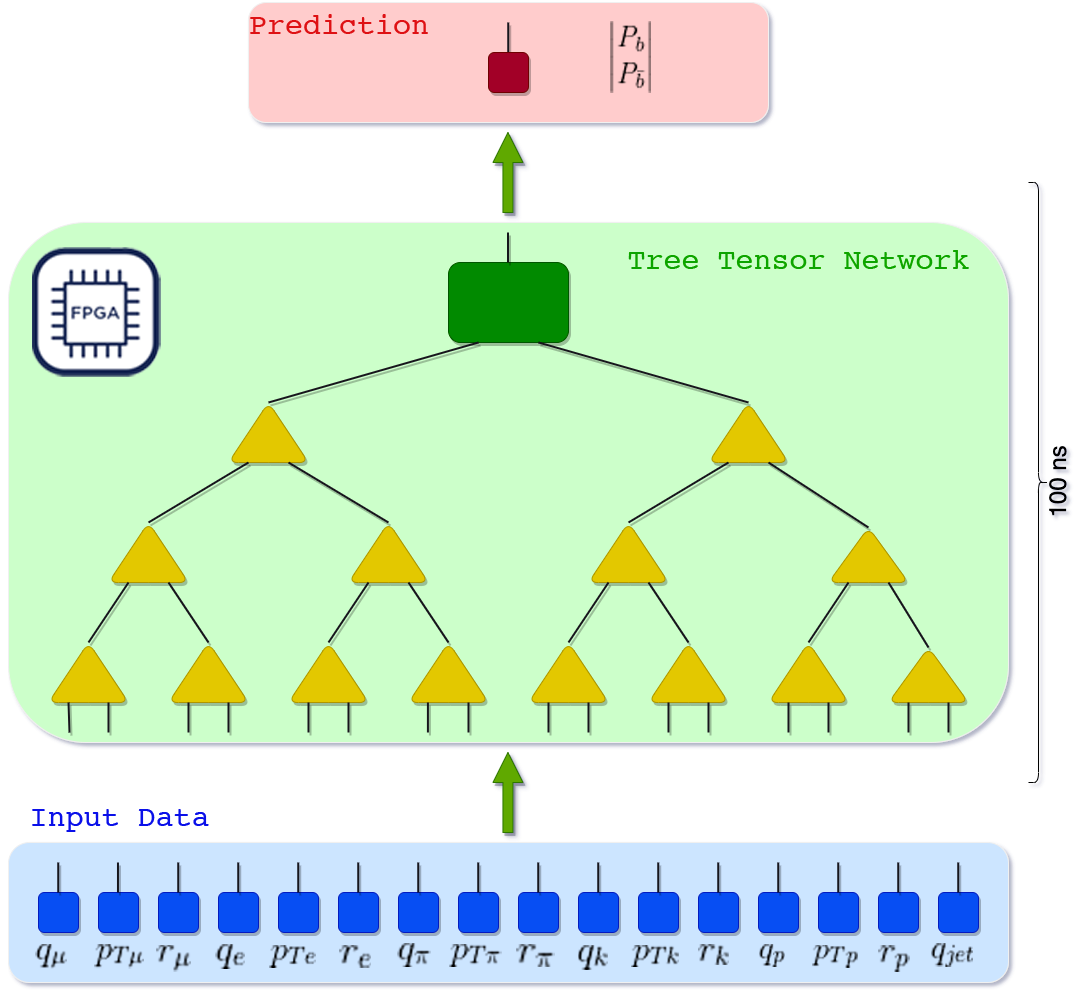}
    \caption{Tree Tensor Network $b/\bar{b}$ jet classifier implemented on FPGA with sub-microsecond prediction latency.}
    \label{fig:catchy}
\end{figure}

Both these attributes make TTNs particularly suitable for pruning, guaranteeing effective preservation of the majority of information used to solve the final ML task, therefore they can be conveniently deployed in frameworks where resource saving is a crucial prerequisite~\cite{Duarte_2018,boutros2024fieldprogrammablegatearrayarchitecture}. Additionally, the exclusively linear operations happening inside these networks contribute to making them highly compatible for the implementation in hardware devices like Field Programmable Gate Arrays (FPGAs). These devices are naturally very versatile and can be exploited to perform a variety of operations with extremely low latency. Moreover, the architecture of FPGAs makes them inherently good at performing fast parallel computations such as matrix multiplications and tensor contraction, which happen to be the only necessary operations for inference with tensor network methods~\cite{tensornetworkssimulatingquantum}.

The combination of quantum-inspired networks and programmable logic allows us to produce a system that can make predictions on input data in an ultra-low latency environment, resulting to be extremely useful for the deployment in the Trigger pipeline of High Energy Physics (HEP) experiments, where quick decisions need to be made in order to filter and collect the relevant physics data. In this paper, starting with the development in FPGA of simple networks used for benchmarking, we end up showing the hardware implementation of the full particle classifier introduced in~\cite{timo_lhcb} (see Fig.\ref{fig:catchy}). The concept of TTNs and their use as binary classifiers is introduced in Sec.~\ref{sec:Arch}, while the methods used for the hardware implementation are described in Sec.~\ref{sec:Methods}. The projections of the necessary resources and the total latency needed to implement TTN architectures with variable hyperparameters are reported in Sec.~\ref{sec:hw_analysis} and eventually, the setup used for the validation of the implemented TTNs is explained in Sec.~\ref{sec:hw_impl}.

\section{Architecture} \label{sec:Arch}
\begin{figure}[h]
    \centering
    \includegraphics[width=\linewidth]{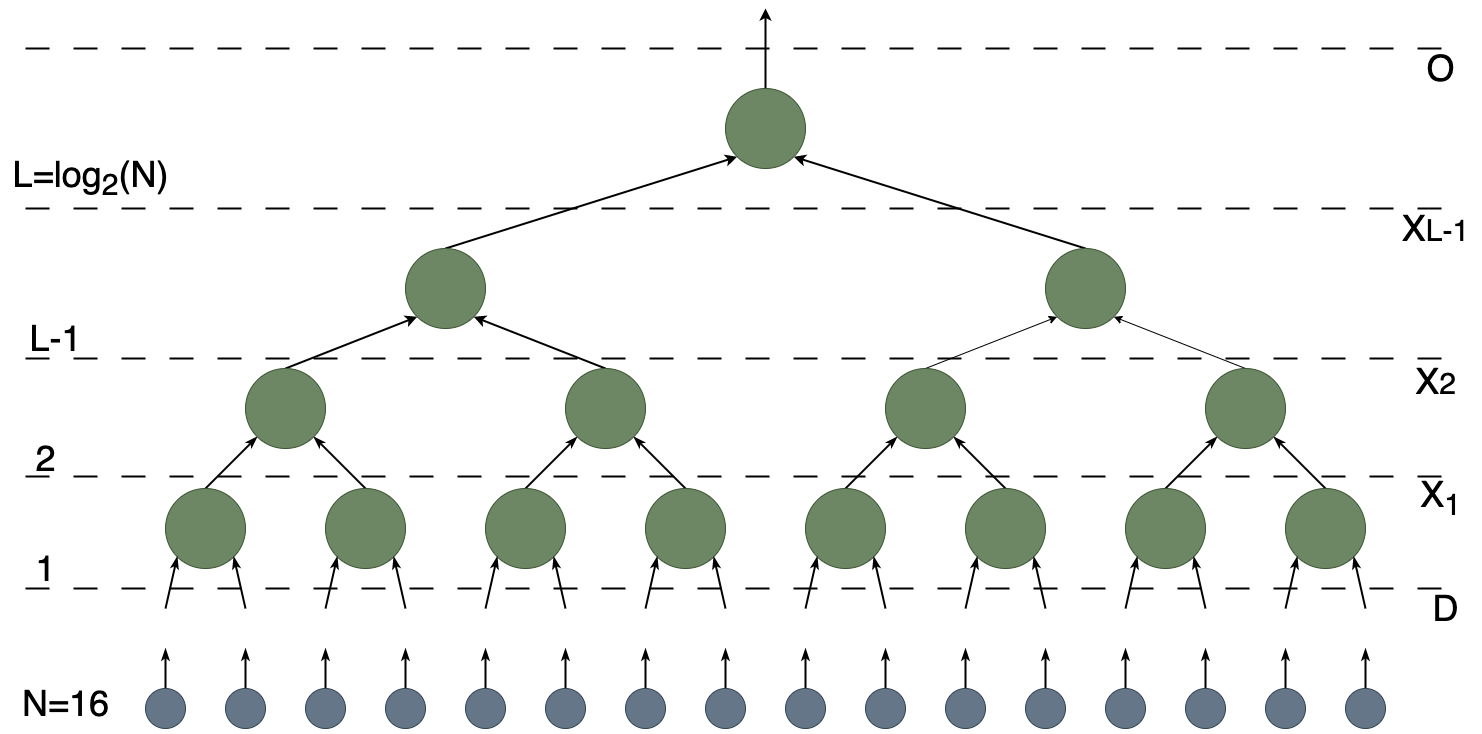}
    \caption{Example architecture with $N=16$ input features and dimensions $\chi_l=[D,\chi_1,\chi_2,\chi_{L-1},O]$.}
    \label{fig:generic}
\end{figure}

Tree Tensor Networks are hierarchical structures made of contracted rank-3 tensors. In general, it is possible to build arbitrary large architectures depending on the number of input features $N$ but for this work, to ease the development in hardware, we limit our study to binary trees, eventually restricting our choices for $N$ only to the powers of 2. Once this parameter is fixed, the number of layers in the tree varies according to $L=\log_2(N)$. The input data are mapped into a higher-dimensional space, following some arbitrary function chosen a priori. Each feature of the dataset assumes the shape of a $D$ dimensional vector (blue tensors in Fig.~\ref{fig:generic}), where $D$ depends on the chosen feature mapping function, therefore representing the data samples as quantum separable states built from the tensor products of the features of the dataset.

 The so-called bond dimension $\chi$ is another fundamental hyperparameter of TTNs, which represents the size of the contracted indices in the inner bonds of the network~\cite{stoudenmire2017}. To make the TTN an exact decomposition of a rank-N tensor this parameter should scale with the layer number $l$ as in $\chi_l=D^{2^{l}}$, nonetheless it can be tuned to reduce the total number of parameters in the TTN and the computational complexity according to $\chi_l=min(D^{2^l},\chi_0)$, where $\chi_0$ is fixed a priori~\cite{Evenbly_2011}. Eventually, the dimension of the output vector $O$ can also be modified depending on the number of classes that need to be identified for the task~\cite{efthymiou2019}. In this work, since we are only performing binary classifications, it will always be either $O=1$ (reducing the output vector to a scalar) or $O=2$, depending on the construction of each network during the trainingphase. In the following, we will identify each architecture with the ordered set of parameters $\chi_l=[D,\chi_1,\chi_2,...,\chi_{L-1},O]$, from which the information on the number of input features can be retrieved by $N=2^L$.

\subsection{Training}

Several models are shown in this work, reporting different combinations of hyperparameters and feature maps. All of them are trained on different datasets to solve increasingly complex tasks and to show the behavior of multiple TTNs, eventually testing the consistency of this study.

A 4-features architecture is trained on the Iris dataset~\cite{Iris}, dividing the total 150 samples in 80\% for the training and 20\% for testing. The input vectors dimension is set to $D=2$ following the spinor-map $f(x)=[\cos{\frac{x\pi}{2}},\sin{\frac{x\pi}{2}}]$, while the bond dimension is instead fixed to $\chi=4$. For this architecture, a final 99\% software classification accuracy was reached and its hardware deployment was done following the Partial Parallel implementation (Sec.~\ref{pp}).

The second TTN is trained on the Titanic dataset~\cite{Titanic}, where the total 850 samples were divided as usual into 80\% for training and 20\% for testing. For this case, the number of input vectors is increased up to $N=8$. Two different mappings are tested, to study any possible change in the predictor's behavior ($f_1(x)=[\cos{\frac{x\pi}{2}},\sin{\frac{x\pi}{2}}]$ and $f_2(x)=[1,x]$). While the input dimension is fixed to $D=2$, we explore different cutting values for the bond dimensions, generating four different models by applying separately $\chi_0=[3,4,8,16]$ to the usual equation $\chi_l=min(D^{2^l},\chi_0)$. The architecture resulting from fixing $\chi_0=8$ and using $f_2$ as mapping achieved the best performance, with an accuracy of \SI{79.3}{\percent} and \SI{74.1}{\percent} on training and test sets respectively; the other networks reached a similar value. The best performing TTN is implemented on FPGA following the Full Parallel implementation (Sec.~\ref{sec:hw_impl}).

Eventually, the 16-features architecture analyzed in~\cite{timo_lhcb} is studied, training it on the LHCb OpenData for $b/\bar{b}$ flavor tagging. The network is built exploiting the spinorial mapping $f_1(x)$ and considering the bond dimension cuts $\chi_0=[8,16]$. It is trained on 400k samples and tested on 80k, reaching a final software classification accuracy of 62\%.

All the above-cited networks are trained with specific minimization techniques, inspired by the "sweeping" algorithm described in~\cite{stoudenmire2017} and adapted to the TTN case. These procedures do not follow classical ML optimization methods (e.g. SGD, ADAM etc.) but they better exploit the full power of tensor network methods by locally updating each tensor, reducing the computational cost and avoiding the learning issue of barren plateau~\cite{barrenplateau}.
To ease the training process and to reduce the probability of falling into local minima, the $\chi_l=[2,4,1]$ and $\chi_l=[2,4,8,1]$ architectures were also initialized following the unsupervised learning technique described in~\cite{unsupervised}.

\subsection{Correlation and Entropy}

A peculiar feature of these TTN-based models is their explainability. As they are the representation of a many-body wave function, we can measure physical quantities of interest, thus interpreting the model and allowing for a clearer understanding of its decision-making process.
In this Section, we concentrate on two measurements: the bipartite entanglement entropy and the two-site correlations between features.

\textbf{Correlations}. In the particular mapping of data described at the beginning of Sec.~\ref{sec:Arch}, each tensor of the resulting many-body quantum state represents a feature (\textit{basis encoding}). Therefore, each site of the quantum state represented by the TTN is also linked to a feature. If we measure the quantum correlations between sites of the TTN we are measuring the correlations between features of the dataset, as learned by the TTN model. To do so, data are encoded through the spinorial map. Consequently, $\sigma^n$ correlations are measured, where $\sigma^n$ is a Pauli operator. \\
As an example, we can measure the two-site spin correlation along the $z$ axis:
\begin{equation}
C_{i,j}^{o} = \frac{\langle\Psi_{TTN}^o|\sigma_i^z\sigma_j^z|\Psi_{TTN}^o\rangle}{\langle\Psi_{TTN}^o|\Psi_{TTN}^o\rangle}
\end{equation}
where $o=1,\dotsc,O$ runs over the output dimension and $\Psi_{TTN}^o$ is the wave function corresponding to a class, obtained by fixing the output index. 
As in statistics, these quantum correlations return $C_{i,j}=1$ if the two $i$ and $j$ spin sites in the TTN many-body state are totally correlated, $C_{i,j}=-1$ if they are anti-correlated, and $C_{i,j}=0$ if there is no correlation between the two.
\begin{figure}[ht]
    \includegraphics[width=\linewidth]{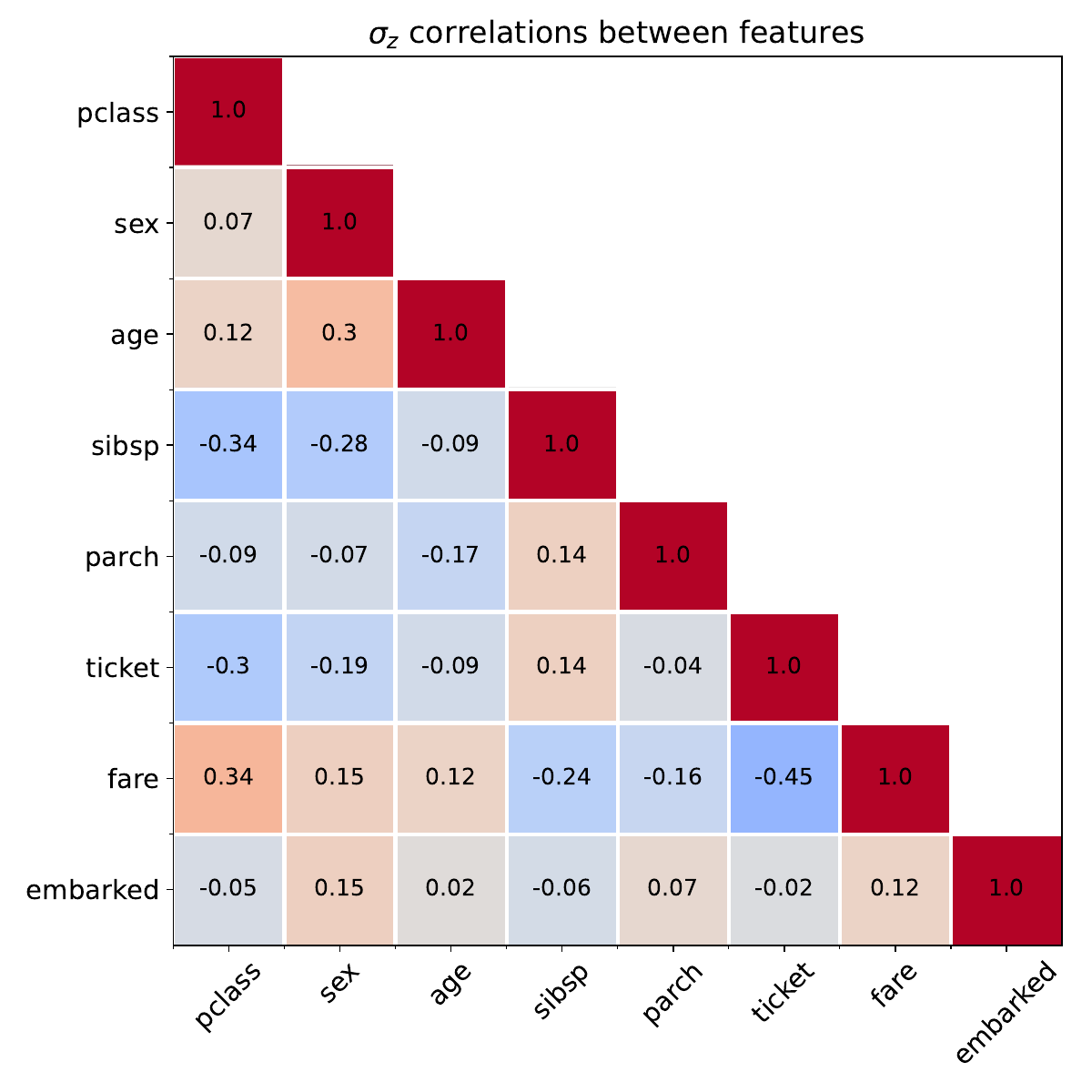}
    \caption{Two-site $\sigma_z$ correlations between features of the titanic dataset as learned by the model.}
    \label{fig:corr}
\end{figure}

\textbf{Entropy}. TTNs are loop-less structures. Therefore, if we cut an internal bond, the whole system $AB$ is divided into two subsystems $A$ and $B$. Then, we can associate each internal bond with the bipartite entanglement entropy between these two subsystems. This can be applied also to the physical links of the TTN, directly connected to the sites of the many-body state. In this case, assuming a basis encoded dataset, we are measuring the bipartite entanglement entropy between a single feature and the rest of the system. \\
This quantity is defined as the von Neumann entropy $S$ of either of the two subsystems, which is
\begin{equation}
    S(\rho_A) = -\operatorname{Tr}[\rho_A\log\rho_A] = -\operatorname{Tr}[\rho_B\log\rho_B] = S(\rho_B)
    \label{eq:vonneumann}
\end{equation}
where $\rho_A=\operatorname{Tr}_B[\rho_{AB}]$, $\rho_B=\operatorname{Tr}_A[\rho_{AB}]$ are reduced density matrices and $\rho_{AB}=|\Psi_{AB}\rangle\langle\Psi_{AB}|$ is the density matrix of the whole system.
In practice, it is inconvenient to calculate explicitly these density matrices; to simplify the calculations, Eq.\ref{eq:vonneumann} can be expressed in terms of the singular values of a Schmidt decomposition of the system. If the TTN is \textit{isometrized} towards one of the two tensors associated with the bond involved, these values can be obtained with a simple Singular Value Decomposition (SVD) on that tensor \cite{timo_lhcb,PhysRevResearch.5.013031}. Note that the amount of entropy is bounded by the size $\chi$ of the bond, $S_{max} = \log(\chi)$.
\begin{figure}[h]
    \centering
    \includegraphics[width=\linewidth]{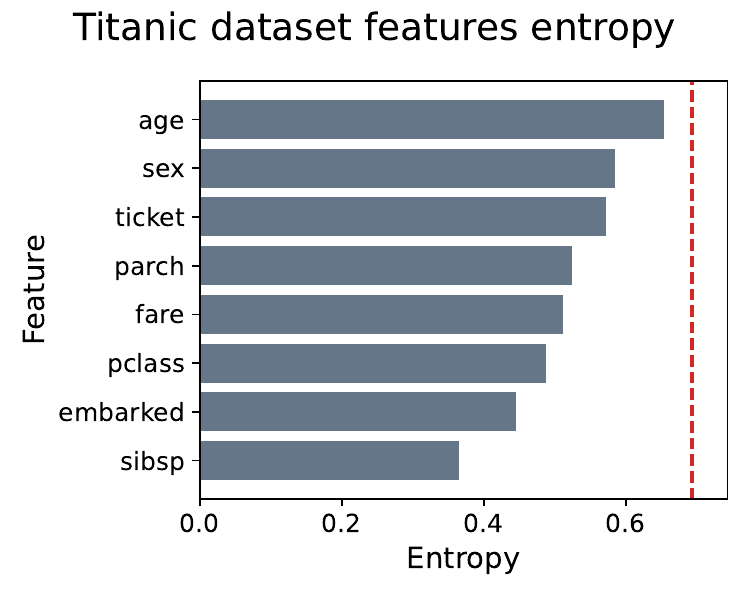}
    \caption{The bipartite entanglement entropy of the features of the titanic dataset. This is measured by cutting the physical bonds of the TTN. The red line is the maximum entropy allowed on those bonds.}
    \label{fig:entropy}
\end{figure}

The results of these measurements help to interpret the model, providing insights on how the information is spread across the network and how the features are combined to produce the output. In practice, they provide a method to rank the features according to their importance for the task, thus allowing the elimination of the least relevant ones and reducing the number of parameters in the network, without a substantial drop in performance. \\
The results depicted in Figures \ref{fig:corr} and \ref{fig:entropy} are measured on the titanic model. As a proof of concept, we trained a model on the four features with the highest entropy. This resulted in a substantial reduction in the number of parameters, from $384$ to only $48$, with an accuracy decrease of a few percent points.
These compression capabilities, outlined in~\cite{timo_lhcb}, are extremely important to fit these models into the limited resources of FPGAs.

\section{Methods} \label{sec:Methods}

\begin{figure}[h]
    \centering
    \includegraphics[width=\linewidth]{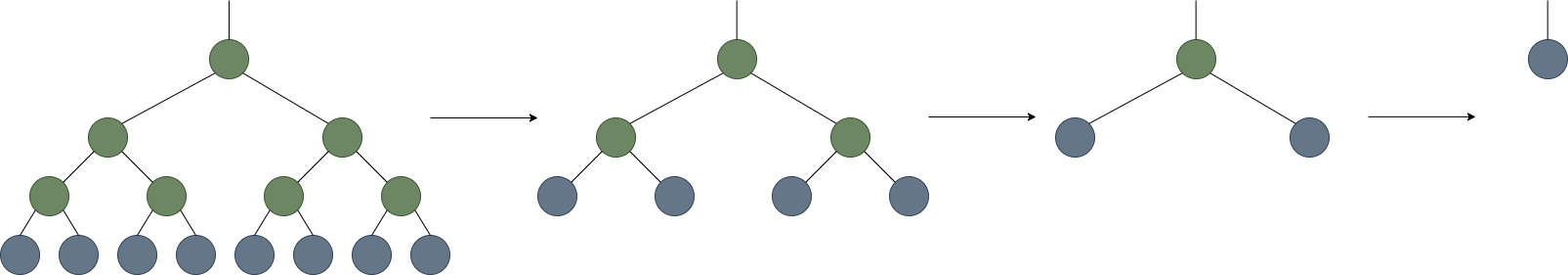}
    \caption{Example of full network contraction for a TTN with $N=8$: the final vector on the right (scalar for $O=1$) is the result of inference.}
    \label{fig:contr_a}
\end{figure}

In this section, a TTN architecture is decomposed in its elementary operations, discussing two strategies for implementing it in hardware, exploiting different degrees of parallelization (Sec.~\ref{fp} and~\ref{pp}). To perform inference on FPGA it means to contract the full TTN architecture (see Fig.~\ref{fig:contr_a}) with the tensors received in input, which represent the data sample that has to be classified. The final output, resulting from tensorial contraction, is a vector (scalar for $O=1$) that encodes the probabilities of each sample to belong to each class.

\subsection{Tensor contraction}
\begin{figure}[h]
    \centering
    \includegraphics[width=0.6\linewidth]{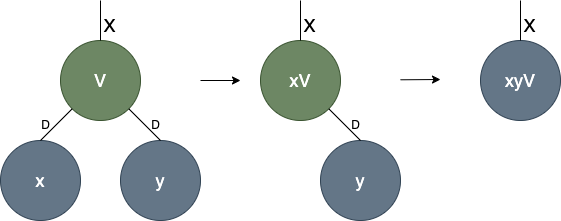}
    \caption{Single node contraction between two vectors of dimension [D] and a tensor of dimension [D,D,X].}
    \label{fig:contr_b}
\end{figure}

To perform inference with TTNs, the fundamental component that needs to be implemented in hardware is the single node contraction. Such operation is represented in Fig.~\ref{fig:contr_b}, considering the example of two D-dimensional feature vectors contracted with a single node of the network; since the result is a vector itself, the contraction of the full architecture must be interpreted as the iteration of said procedure for all the layers of the TTN.

The contraction operation is algebraically computed by the following equation: 
\begin{equation}
    z_i=\sum_j\sum_kx_jy_kV_{ijk}
    \label{eq:contr}
\end{equation}
where $x$ and $y$ are the input vectors and $V$ is a rank-three tensor of the TTN. To do arithmetics on FPGA we make use of Digital Signal Processors (DSPs), which can compute the result of two-factor multiplications, together with many other more complex operations~\cite{9040130}. As a consequence, to meet the physical limits of DSPs, the three-factor multiplication of the tensor contraction must be split into two stages, concatenating the results of two multipliers (DSPs) and eventually summing them together. On this implementation in particular, we first compute the cartesian product between the two input vectors $x$ and $y$, and eventually multiply the result with the values of the weights in the rank-three tensors $V$ of the network. 

To explore different degrees of parallelization for this operation, two separate implementations are tested. The Full Parallel (FP) approach is built to minimize the latency while maximizing the usage of DSPs, therefore performing all the calculations in parallel. On the other hand, a Partial Parallel (PP) implementation is conceived to reduce the total number of DSPs used for a single contraction, paying the price of this saving with a consequential increase in the overall latency of the algorithm.

These implementations are not unique and do not claim to be optimal: many other configurations and trade-offs between resources and latency could be implemented, even if they are not explored in this work. For example, the number of DSPs exploited in the PP implementation could be further reduced, leading to a completely Non-Parallel (NP) computation, useful to be deployed in environments with an extremely limited number of resources. At the same time, multiplications could also be forced to be executed with the FPGA look-up tables (LUTs), provided that the numerical values in hardware are represented with small vectors of bits, moving the overall resources optimization to a completely different space (Sec.~\ref{sec:quantization}).

\subsection{Full Parallel}\label{fp}

Within the FP implementation, one DSP is devoted to each single two-factor multiplication happening in the tree. In this way, there is no reuse of resources and all the calculations can be brought on parallelly. The computation of the cartesian product between $x$ and $y$ is performed at the first stage of multiplication; once these results are computed, they are forwarded to the second layer of DSPs and multiplied by the node weights. Eventually, the values of the three-factor products are all summed together by exploiting Adder Trees (ATs), the implementation of which involves no DSPs.

\begin{figure}[h]
    \centering
    \includegraphics[width=\linewidth]{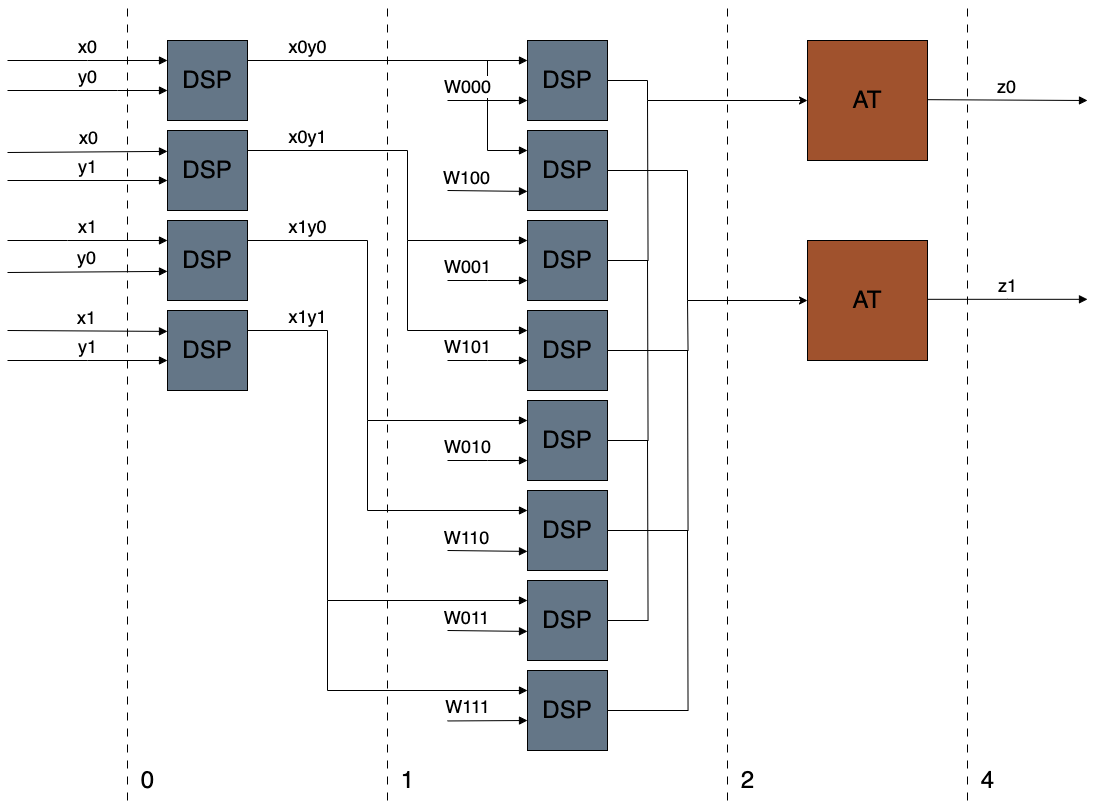}
    \caption{Example of FP implementation block diagram, considering $D=2$ and $\chi=2$. Parallel computation timings are reported in $\Delta t_{DSP}$ units.}
    \label{fig:fp}
\end{figure}

The block diagram in Fig.~\ref{fig:fp} reports the graphical representation of said procedure for the example case of $D=2$, $\chi=2$. From this, it is easy to understand how the number of resources varies according to the different parameters of the node: the first multiplication stage always requires $D^2$ DSPs to compute all the possible products between the components of $x$ and $y$ vectors. Moreover, a set of $\chi$ DSPs needs to be used for every $D^2$ value, each one corresponding to a single weight, resulting in a total amount of $\chi D^2$ resources deployed for the second stage of multiplication. The number of values to be summed together, corresponding to the inputs of the adder trees, is therefore fixed at $D^2$, while the number of necessary adder trees is always constant to the length $\chi$ of the output vector.

\begin{figure}[h]
    \centering
    \includegraphics[width=\linewidth]{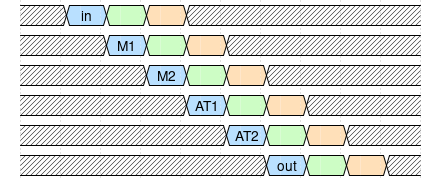}
    \caption{Wave diagram for FP implementation considering $D=2,\chi=2$, timing in $\Delta t_{DSP}$ units. Different colors represent different data samples. M1/2 refers to the two stages of multiplication, AT1/2 corresponds to the layers of each AT.}
    \label{fig:fp_wave}
\end{figure}

Each multiplication stage in the tensor contraction can take from 1 to 4 clock cycles, depending on the number of registers used within the DSPs and parametrized with $\Delta t_{DSP}$. For the FP implementation, all computations are performed in parallel, moving the overall latency of the algorithm to its absolute minimum. In particular, every multiplication step always requires $\Delta t_{DSP}$ clock cycles to be completed, allowing the results to be available all at the same time and enabling the usage of adder trees, which naturally require synchronous input values. In the end, computations for different input samples are properly pipelined as reported in Fig.~\ref{fig:fp_wave}.

\subsection{Partial Parallel}\label{pp}

\begin{figure}[h]
    \centering
    \includegraphics[width=\linewidth]{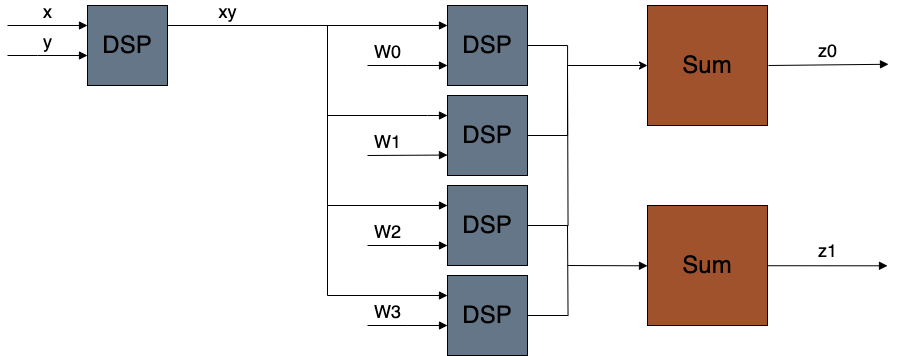}
    \caption{Example of PP implementation block diagram, considering $D=2$ and $\chi=2$. }
    \label{fig:pp_contr}
\end{figure}

\begin{figure}
    \centering
    \includegraphics[width=\linewidth]{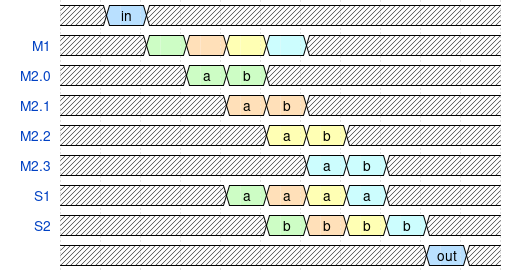}
    \caption{Wave diagram for PP implementation considering $D=2,\chi=2$, timing in $\Delta t_{DSP}$ units. Different colors represent different data samples. M1 refers to the first stage of multiplication, while the M2.i lines show the $D^2$ DSPs at the second layer of multiplication. S1/2 represent the $\chi$ serial sums.}
    \label{fig:pp_wave}
\end{figure}

In the PP case instead, only one DSP is devoted to the computation of the cartesian product between $x$ and $y$ at the first stage of multiplication. All the possible combinations of their components are sent in input to the DSP at different times, and the results are returned in output accordingly. Once they are registered, they are read by the $D^2$ DSPs present at the second stage of multiplication. This layer is also performing its computation serially, since different weights are read at different clocks' rising edges, scanning all the $\chi$ values necessary to produce the three-factor products (Fig.~\ref{fig:pp_wave}). Eventually, since the computation is not parallel, ATs are not used for the final sum; in this case, all the values can be added together sequentially in an accumulator right after they become available.

This approach is devised to reduce the total amount of DSPs in use for a single contraction, even if it naturally causes an increase in the overall latency of the operation. From the block diagram in Fig.\ref{fig:pp_contr} it is clear how this implementation always requires only $(D^2 + 1)$ DSPs, completely removing the dependence of the number of resources on the length of the output vector $\chi$, which can become considerably large in some TTN architectures. This configuration can therefore represent a convenient solution for the implementations of TTN in systems in which the latency requirements are not too strict but in which the limitation of available resources is more constraining.

\section{Hardware analysis} \label{sec:hw_analysis}

In this section, the scalings of resources and latency of the whole TTN architectures are reported, following the two tensor contraction implementations presented in Sec.~\ref{sec:Methods}. Eventually, quantization studies are shown, investigating how different choices for numeric representation can affect the performances of inference in hardware.

\subsection{Resources}

The FP and PP implementations allow us to derive a precise calculation for the necessary resources of a single node contraction, therefore the projection of said calculation for a full TTN architecture is completely deterministic. It is straightforward to compute the scaling of the total number of DSPs with respect to the main TTN parameters N, D, $\chi$ and O.

Considering the FP implementation, each node requires $\chi^2_{l-1}(\chi_{l}+1)$ DSPs, where $\chi_{l-1}$ and $\chi_l$ represent respectively the input and output vector lengths at every layer. For the PP approach instead, there are always $(\chi^2_{l-1}+1)$ DSPs involved in every node. For binary trees, the number of nodes per layer is equal to $\frac{N}{2^l}$, where $l$ naturally scales from 1 to $L=\log_2(N)$. Putting together all this information, we derive Eq.\ref{eq:dsp_fp}-\ref{eq:dsp_pp}, considering the notation $\chi_i=[D,\chi_1,\chi_2,...,\chi_{L-1},O]$ for enumerating the contracted dimensions of tensors involved in the network.

\begin{figure}[h]
    \centering
    \includegraphics[width=\linewidth]{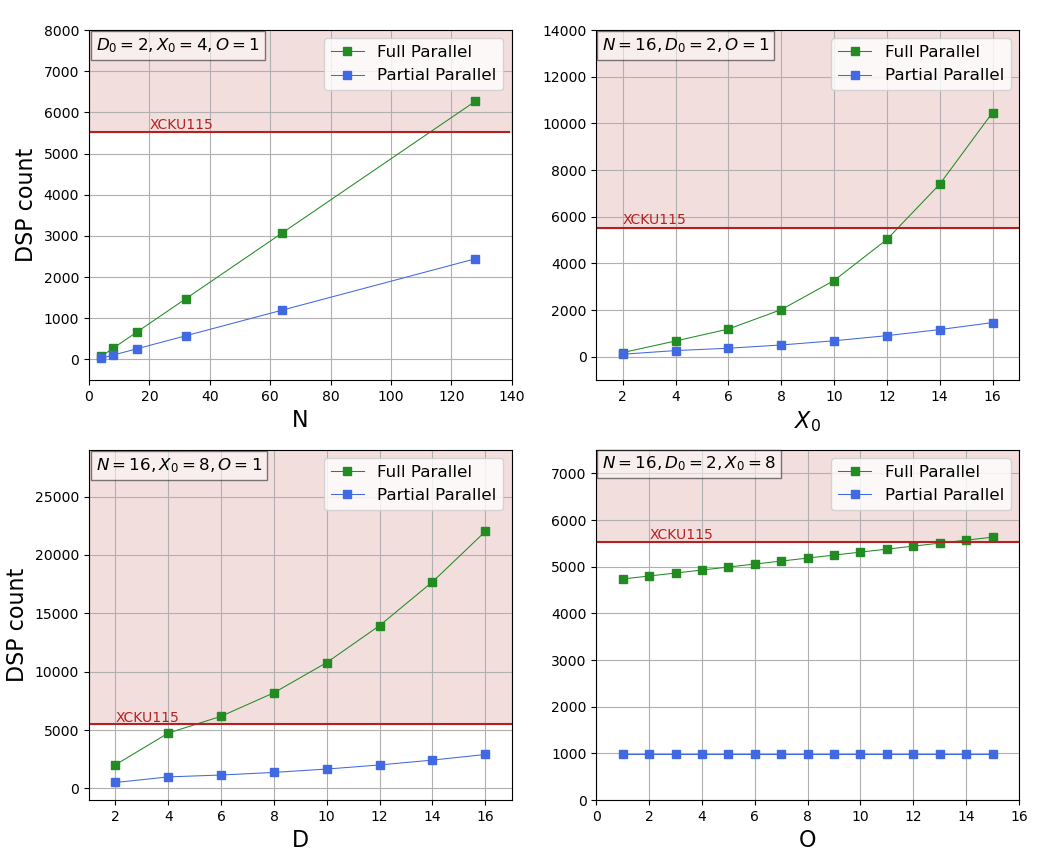}
    \caption{DSP number projections of FP and PP implementations for different combinations of input features N, feature map dimension D, bond dimension $\chi_0$ and output classes O.}\label{fig:resource_projection}
\end{figure}

\begin{equation}
    DSP_{FP}=\sum_{l=1}^{L}\chi_{l-1}^2(\chi_l+1)\frac{N}{2^l}
    \label{eq:dsp_fp}
\end{equation}
\begin{equation}
    DSP_{PP}=\sum_{l=1}^{L}(\chi_{l-1}^2+1)\frac{N}{2^l} 
    \label{eq:dsp_pp}
\end{equation}

The plots in Fig.~\ref{fig:resource_projection} report the total amount of DSPs for various combinations of N, $\chi$, D and O, according to what is computed by Eq.~\ref{eq:dsp_fp}-\ref{eq:dsp_pp} for the FP and PP cases, together with the limit of resources fixed by the target FPGA (XCKU115) chosen for the first firmware implementation. The overall amount of DSPs scales linearly with $N$ for both FP and PP implementations, while it shows a polynomial growth for D and X, where the discontinuity in their behavior can be interpreted as the switch of minimum value according to $\chi_l=min(D^{2^l},\chi_0)$. Instead, regarding the classification output $O$, the scaling is linear for the FP case, while the number of DSP does not depend on the number of classes of the problem in the PP approach, due to the sequential nature of the final summing in the contraction of the full TTN.

\subsection{Latency}

The two implementations also report a different behavior in terms of latency. Since the FP approach maximizes the usage of resources, aiming at minimizing the time of the computation, it reaches the limit in which the total latency of the TTN scales logarithmically with the input vector dimension of each layer $\chi_{l-1}$. The PP case instead suffers from an increase in latency, due to the reuse of the same resources for different computations, leading to a quadratic scaling with the input dimension of each node. Eq.~\ref{eq:lat_fp}-\ref{eq:lat_pp} allow us to compute the number of clock cycles needed to perform a complete contraction of a TTN architecture with parameters $\chi_i=[D,\chi_1,\chi_2,...,\chi_{L-1},O]$.

\begin{equation}
    LAT_{FP}=\Delta t_{DSP}\sum_{l=1}^{L}2 + \log_2(\chi_{l-1}^2)
    \label{eq:lat_fp}
\end{equation}

\begin{equation}
    LAT_{PP}=\Delta t_{DSP}\sum_{l=1}^L\chi_{l-1}^2+\chi_l+1\quad\text{if}\quad \chi_{l-1}\le \chi_l
     \label{eq:lat_pp}
\end{equation}

For the FP scenario, the time necessary for the contraction of each layer depends logarithmically on the square of the length of the input vectors. Since the multiplications are performed parallelly in this case, any change in the overall latency is due to the presence of ATs, which will always have $\chi_{l-1}^2$ values to be summed together, generating $\log_2(\chi_{l-1}^2)$ summing layers, each one needing $\Delta t_{DSP}$ clock cycles to perform its computation. In the PP case instead, the latency estimation is slightly more complicated due to the serial distribution of each result in time. The variable $\Delta t_{DSP}$ is inserted in both cases to parametrize the number of clock cycles needed for a single DSP to perform a multiplication, which of course can vary in different implementations of the firmware, depending on the clock frequency in use and on the latency requirements of the systems. In our case for simplicity, it has been fixed to $\Delta t_{DSP}=1$ for all implementations.

\begin{figure}[h]
    \centering
    \includegraphics[width=\linewidth]{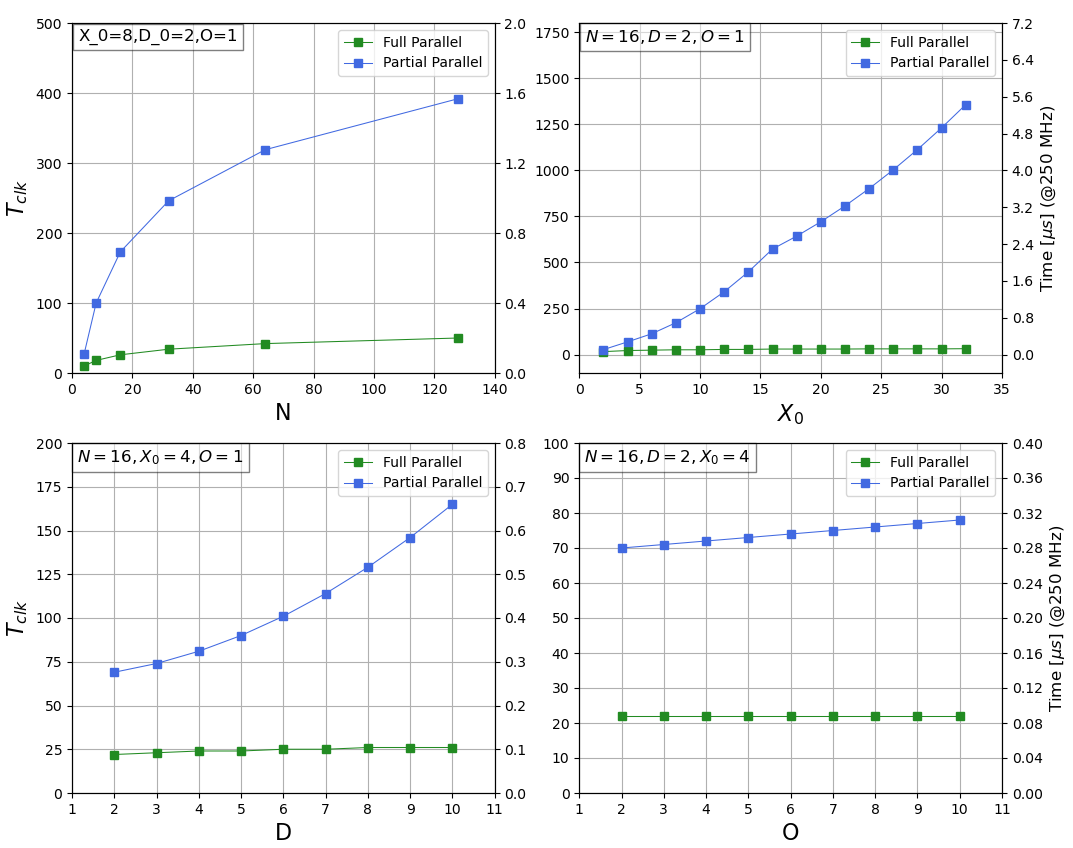}
    \caption{Latency projection of FP and PP implementations for different combinations of input features N, feature map dimension D, bond dimension $\chi$ and output classes O. Right axes show $\mu s$ values considering a 250 MHz clock.}\label{fig:latency_projection}
\end{figure}

The plots reported in Fig.~\ref{fig:latency_projection} show how the latency varies with the different hyperparameters of the networks, measuring such value both in terms of clock cycles (left axis) and absolute time (right axis). For this computation, a period of $T_{clk}=4 ns$ is considered, corresponding to the 250 MHz clock used in our implementations. 

The total latency of the networks scales logarithmically with the number of input features N for both the FP and PP implementations, even if they are multiplied by very different factors depending on the other parameters of the network. Considering the dependence on D and X instead, the FP case reports a logarithmic behavior due to the presence of ATs in its implementation, while the PP approach shows a quadratic scaling with respect to both parameters. Even in this case, the sudden change of behavior in the bond dimension plot must be interpreted as the action of the minimum equation $\chi_l=min(D^{2^l},\chi_0)$. Eventually, the latency for the FP networks does not depend on the number of classes involved in the problem, since the final summing is performed completely in parallel by the ATs. The PP case shows the usual linear behavior with respect to the parameter O.

\subsection{Quantization}\label{sec:quantization}

All the numbers involved in the networks are real values belonging to the range [-2,2], according to the normalization applied in the software processing of the TTNs. A priori, these are inserted in firmware exploiting a total of 16 bits, filled according to their fixed-point representation. Nonetheless, different TTN architectures might require different numeric precision, therefore individual quantization studies are necessary.

By reducing the total amount of bits used to represent each number in firmware, one can obtain a consistent gain in terms of resources in use. In particular, a single two-factor multiplication can be implemented with different combinations of Flip Flops (FF), Look Up Tables (LUT), or Digital Signal Processors (DSP), depending on the length of the logic vectors used to represent each real value. If a specific network does not require excessive numeric precision for its hardware implementation, it is possible to minimize the number of bits used to represent each value in the TTN without losing any classification power, therefore possibly avoiding the direct usage of DSPs to perform the tensorial contractions.

\begin{figure}[h]
     \centering
     \includegraphics[width=\linewidth]{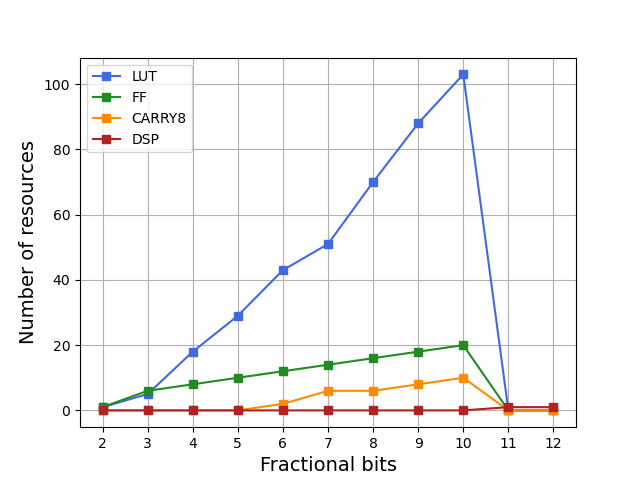}
   \caption{Logic resources needed on XCKU115 for a two-factor multiplication between 16b vectors, considering a variable number of fractional bits.}\label{fig:singlemult}
\end{figure}

Fig.~\ref{fig:singlemult} shows the amount of resources used to implement a single two-factor multiplication on XCKU115. These values are obtained considering 2 bits for the sign and integer part of the two factors and by varying the number of bits in their fractional parts, from 2 to 12. In this case, the multiplication is performed with FFs and LUTs until the number of fractional bits is below 10: above this threshold, the hardware synthesizer substitutes all the logic involved within a single DSP. These results are hardware-specific since they depend on the total amount of resources available on the FPGA in use. Nonetheless, they are useful for understanding how to avoid the direct usage of DSPs in our specific implementation.

\begin{figure}[h]
    \centering
    \includegraphics[width=\linewidth]{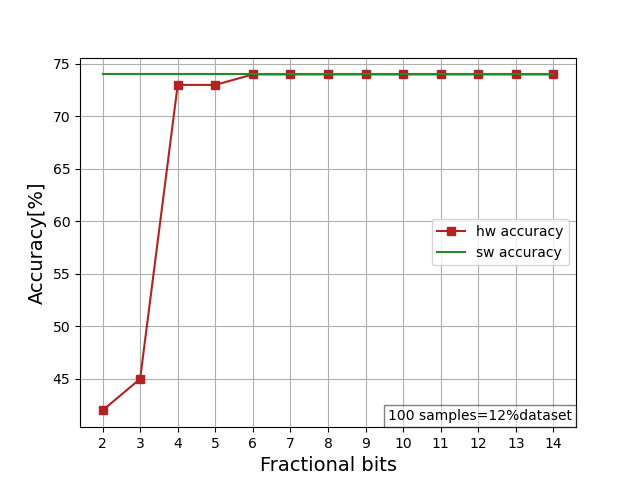}
    \caption{Hardware classification accuracy with respect to the number of fractional bits. Values computed for the [2,4,8,1] TTN, tested on 12\% of the Titanic dataset.}\label{fig:quanttitanic}
\end{figure}

Fig.~\ref{fig:quanttitanic} reports the quantization study performed on the [2,4,8,1] TTN trained on the Titanic dataset. For this particular tree, the number of fractional bits can be reduced from 14 to 6 without causing any loss in the classification accuracy of the network. As a consequence, this architecture can be implemented in hardware exploiting fewer DSPs compared to what is expected by Eq.~\ref{eq:dsp_fp}-\ref{eq:dsp_pp}. With this in mind, it is possible to think of said equations as upper-bound estimation of the number of resources required for each TTN implementation, considering that further optimization might be possible after having performed quantization studies. 

\section{Hardware implementation} \label{sec:hw_impl}

The main goal of this work is to produce a network that can be deployed on FPGA to perform binary classification with ultra-low latency. To accomplish this, TTN architectures are firstly trained in software and then, once the target accuracy is reached, the inference is offloaded in hardware. 

The topology of the architecture is chosen a priori, programming the FPGA with a firmware that contains the fixed hyperparameters of the network. The trained values contained in each tensor are loaded in memory blocks (BRAM), in this way easing the test of multiple networks with the same set of hyperparameters. The logic is endowed with a series of registers that can be read and written by the host PC via AXI Lite protocol~\cite{axilite}, and that are mapped to each single weight in the TTN. The inference process itself is performed by sending a stream of input data and collecting the corresponding output values produced by the TTN, exploiting the AXI Stream protocol~\cite{axistream} for the Host/FPGA communication. The whole process is validated by comparing said values to the results obtained in software for the same architecture and the same subset of input data.

\begin{figure}[h]
    \centering
    \includegraphics[width=\linewidth]{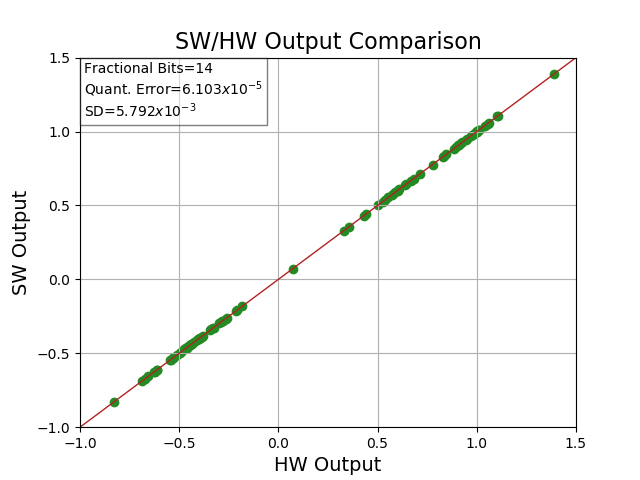}
    \caption{Software and hardware output values comparison for Titanic [2,4,8,1] over 100 data samples, considering the maximum numeric precision of 14 fractional bits.}
    \label{fig:outputval}
\end{figure}

The output comparison in Fig.\ref{fig:outputval} corresponds to the Titanic [2,4,4,1] architecture implemented with the FP approach, where the numbers have been represented with the maximum fixed point precision of 14 bits for the fractional part, corresponding to a quantization error of $6.103\cdot10^{-5}$. The hardware output values match those obtained in the software, with a standard deviation of $5.792\cdot10^{-3}$; this result is enough to guarantee a 1-to-1 match with the software and hardware classification labels.

\subsection{LHCb predictor}
In this paragraph, the results for the hardware implementation of the 16-features TTN trained on LHCb open data and described in~\cite{timo_lhcb} are reported. To facilitate the training, the tree was top-isometrized (see Fig.\ref{fig:catchy}), eventually implementing in hardware the normalized version of the top tensor. For this network, the output is a vector that stores the probabilities of belonging to each class ($b/\bar{b}$), therefore producing results in the range [0,1].

\begin{figure}[h]
    \centering
    \includegraphics[width=\linewidth]{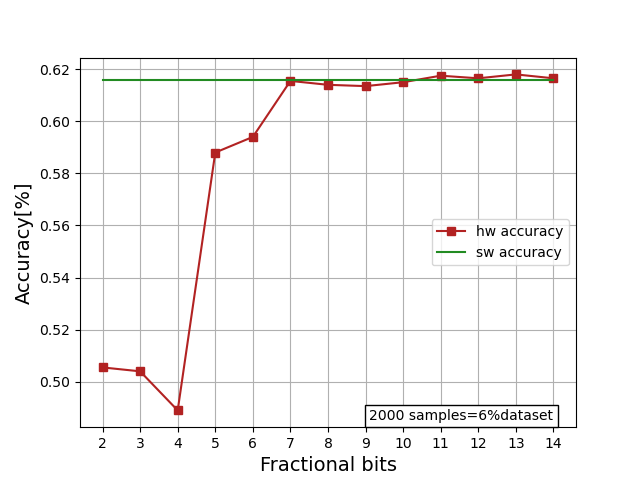}
    \caption{Hardware classification accuracy with respect to the number of fractional bits. Values computed for the [2,4,8,8,2] TTN, tested on 6\% of the LHCb dataset~\cite{timo_lhcb}.}\label{fig:quantlhcb}
\end{figure}

The same quantization studies performed for the Titanic [2,4,8,1] network were reproduced for the LHCb [2,4,8,8,2] tree, as can be seen in Fig.\ref{fig:quantlhcb}, analyzing the inference results for 2000 data samples. In this case, the hardware accuracy matches its software counterpart only when the number of bits devoted to the fractional part remains greater than 7; below this value, the network cannot guarantee a reliable predicting behavior. This result confirms that the choice for numeric precision is architecture and task-specific.

\begin{figure}[h]
    \centering
    \includegraphics[width=\linewidth]{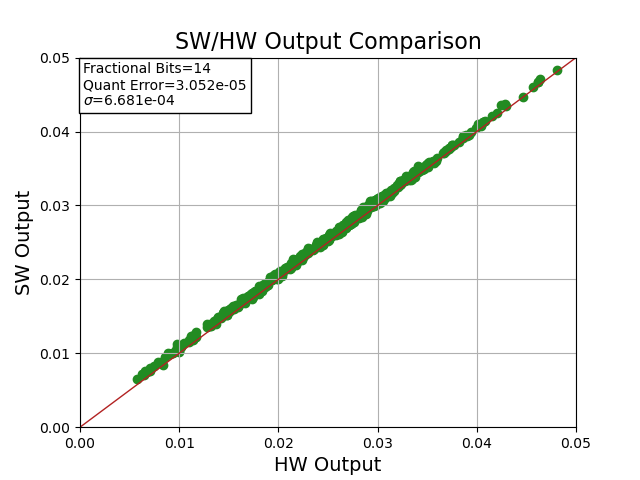}
    \caption{Software and hardware normalized output values comparison for LHCb [2,4,8,8,2] over 500 data samples, considering the maximum numeric precision of 14 fractional bits.}
    \label{fig:outputlhcb}
\end{figure}

The plot in Fig.~\ref{fig:outputlhcb} shows the comparison of hardware and software results of inference for a total of 500 data samples. Since the hardware values are derived from the normalized top tensor, the comparison is performed by dividing the software outputs by its norm, therefore restricting the results to the [0,0.05] range.  With the usual maximum quantization precision of 14 fractional bits, the output values are compatible up to a squared mean difference of $6.681\cdot10^{-4}$. Even in this case, the hardware classification accuracy matches completely its software counterpart.

\section{Conclusion} \label{sec:Conclusion}
In this work, several TTN architectures for binary classification are presented, solving common ML tasks as well as harder classification problems coming from physics datasets. Their quantum characteristics are explored to investigate the distribution of the learned information, with the aim of compressing said networks for optimized hardware deployment.

The logic of the VHDL firmware used for their implementation is explained, focusing on two different degrees of parallelization used to perform the inference on FPGA. A deterministic projection for the values of resources and latency is also reported, highlighting how the PP and FP approaches can allow the deployment of a wide range of TTN topologies. Quantization studies are performed, providing an additional method for further compressing TTNs, in this way easing their hardware implementation by optimally tuning the usage of resources with the numeric precision required by each classification task.

\begin{table}[h]
\centering
\resizebox{\columnwidth}{!}{%
\begin{tabular}{|c|c|c|c|c|c|} 
 \hline
 Dataset & TTN  & DSP & BRAM &  Latency\\ 
 \hline\hline
 Iris & [2,4,1] PP & 1\% & 2\% & 108 ns\\ 
 \hline
 Titanic & [2,4,8,1] FP & 8\% & 19\% & 72 ns\\
 \hline
 LHCb & [2,4,8,8,1] FP & 36.5\% & 84\% & 104 ns\\
 \hline
\end{tabular}%
}
    \caption{Firmware occupancies and final latency for different TTNs. Values calculated considering the XCKU115 implementation with $T_{clk}=250MHz$ and$\Delta t_{DSP}=1$}
    \label{tab:recap_table}
\end{table}

In conclusion, the inference algorithm in hardware is validated and compared with software, allowing the TTN prediction to be exactly reproduced on FPGA. The LHCb tree described in ~\cite{timo_lhcb} is successfully offloaded in hardware, showing a sub-microsecond inference behavior and probing the possibility of deploying this type of networks in the trigger pipeline of HEP experiments (see Tab.\ref{tab:recap_table}).

\bibliographystyle{unsrt}  

\vfill

\end{document}